\documentclass[epj]{svjour}
\usepackage{graphics}
\usepackage{color}
\usepackage[dvips]{graphicx}
\usepackage{dcolumn}
\usepackage{bm}
\begin{document}
\title{Precision measurement of the branching fractions of the 4p~$^2$P$_{3/2}$ decay of Ca~II}
\author{R.~Gerritsma\inst{1} \and G.~Kirchmair\inst{1,2} \and F.~Z{\"a}hringer\inst{1,2} \and J.~Benhelm\inst{1,2} \and R.~Blatt\inst{1,2} \and C.~F.~Roos\inst{1,2}
}                     
%
%
\institute{Institut f{\"u}r Quantenoptik und Quanteninformation,
{\"O}sterreichische Akademie der Wissenschaften,
Otto-Hittmair-Platz 1, A-6020 Innsbruck, Austria \and Institut
f{\"u}r Experimentalphysik, Universit{\"a}t Innsbruck,
Technikerstra{\ss}e 25, A-6020 Innsbruck, Austria\\
\email{Christian.Roos@uibk.ac.at}}
\date{Received: \today}
%
\abstract{ We perform precision measurements of the branching
ratios of the 4p~$^2$P$_{3/2}$ level decay of a single
$^{40}$Ca$^+$ ion suspended in a linear Paul trap. High precision
is achieved by a novel technique based on monitoring the
population transfer when repeatedly pumping the ion between
different internal states. The branching fractions into the
4s~$^2$S$_{1/2}$, 3d~$^2$D$_{5/2}$ and 3d~$^2$D$_{3/2}$ levels are
found to be 0.9347(3), 0.0587(2) and 0.00661(4), respectively. For
the branching ratio $A($P$_{3/2}-$S$_{1/2})/\sum_J
A($P$_{3/2}-$D$_J)=14.31(5)$, we find a forty-fold improvement in
accuracy as compared to the best previous measurement.
} 
\maketitle
\section{Introduction}
\label{intro}

A precise knowledge of the radiative properties of stellar matter
\cite{Seaton:1994} is crucial for matching theoretical models to
observations in many domains of astrophysical research
\cite{Rogers:1994}. Modelling isotopic abundances or the energy
transport by photons in a star or a gas, in turn, requires a
precise knowledge of atomic transition frequencies and oscillator
strengths based on atomic structure calculations and experiments
with a wide variety of neutral atoms and ions. Singly charged
calcium ions have been used in various astrophysical observations
\cite{Wolf:2008,Biemont:1996}. In particular, monitoring of
emission and absorption lines of dipole transitions between
low-lying states has provided information about systems like
galaxies \cite{Persson:1988,Nelson:1995}, interstellar gas clouds
\cite{Welty:1996}, gas disks surrounding stars
\cite{Ferlet:1987,Hobbs:1988}, and stars \cite{Mashonkina:2007}.

In laboratory experiments, cold trapped ions have received
considerable attention in optical frequency
metrology~\cite{Diddams:2001}, precision
measurements~\cite{Fortson:1993,Haeffner:2000}, and as a physical
implementation of quantum information
processing~\cite{Cirac:1995}. The ion species Ca$^+$ is of
particular interest in this research, as it has been used for
demonstrating important experimental steps in quantum information
processing~\cite{Riebe:2004,Benhelm:2008}. It is also studied for
use in ion clocks~\cite{Champenois:2004,Chwalla:2008} and for
precision spectroscopy, for instance for the search of possible
drifts of physical constants~\cite{Dzuba:1999,Wolf:2008}.

Theoretical physicists have devoted a considerable effort to
improve calculations of matrix elements of transition rates and
polarizabilities~\cite{Guet:1991,Liaw:1995,Mitroy:2007,Arora:2007,Sahoo:2007}
in Ca$^+$. Singly charged calcium is a quasi-hydrogenic system
that constitutes an interesting model system for testing atomic
structure calculations of other singly charged alkali-earth ions
with higher nuclear charge. An improved knowledge of their atomic
structure would also be of interest for parity non-conversation
experiments \cite{Fortson:1993} in Ba$^+$ or Ra$^+$ that require
precise values of transition matrix elements for a determination
of the strength of parity-violating interactions. For a comparison
of theoretical predictions with experimental observations,
theorists often turn to precision measurements of excited state
lifetimes~\cite{Smith:1966,Gallagher:1967,Gosselin:1988,Jin:1993,Barton:2000,Kreuter:2005}.
Alternatively, precise measurements of branching fractions
\cite{Huber:1986} of the decay of an excited state into
lower-lying states could be used.

Measurements of branching ratios or oscillator strengths in ions
often date back quite a long time~\cite{Gallagher:1967} and are
usually performed on ensembles of ions in discharges, ion beams
\cite{Cox:1999} or trapped clouds of ions. A notable exception is
a recent experiment~\cite{Kurz:2008} measuring the branching
fractions of the P$_{3/2}$ state decay with a single trapped
Ba$^+$. Single trapped and laser-cooled ions form an attractive
system to perform such precision measurements, as state
preparation and quantum state detection can be carried out with
very high fidelity. In addition, the use of single ions eliminates
errors due to depolarizing collisions.

\begin{figure}
\includegraphics[width=68mm]{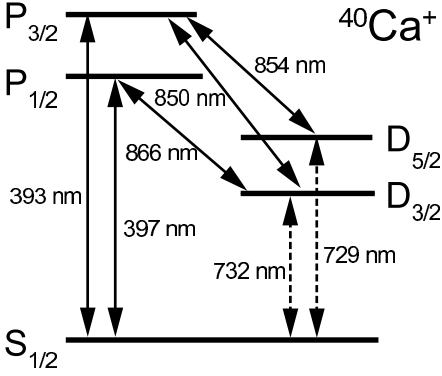}   
\caption{ Level scheme of Ca$^+$ showing its five lowest energy
levels and the transition wavelength of the electric-dipole (solid
lines) and electric-quadrupole transitions (dashed lines). The
metastable D-states have a lifetime of about 1s. In the
experiments, narrow-band laser sources are available to excite all
but the transition at 732 nm.} \label{fig_levelscheme_Ca40}
\end{figure}

For the precision measurements reported in this paper, a single
Ca$^+$ ion held in a linear Paul trap is employed for determining
the branching fractions of the decay of the excited state
4p~$^2$P$_{3/2}$ into the states 4s~$^2$S$_{1/2}$,
3d~$^2$D$_{5/2}$ and 3d~$^2$D$_{3/2}$
(Fig.~\ref{fig_levelscheme_Ca40}). We introduce a novel technique
based on repetitive optical pumping to shuffle populations between
two of the lower-lying states. The shuffling steps are followed by
detection of the population transferred to the third state. This
technique surpasses the branching ratio measurement of
Gallagher~\cite{Gallagher:1967} by a factor of forty in precision
and provides branching fractions with a precision of better than
1\%. In combination with precision measurements of the excited
P$_{3/2}$ state lifetime, our measurements lead to a more precise
determination of the transition probabilities on the
S$_{1/2}$--P$_{3/2}$, D$_{3/2}$-P$_{3/2}$, and D$_{5/2}$-P$_{3/2}$
transitions.

\section{Measurement method}
\label{sec:methods}
\begin{figure}
\includegraphics[width=80mm]{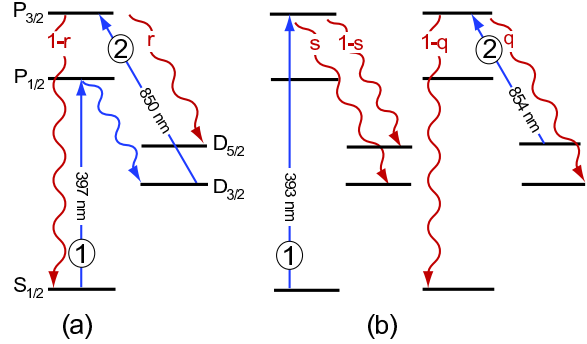} 
\caption{Scheme for measuring the branching fractions of the
P$_{3/2}$ state's decay. (a) To determine the branching ratio $r$
between decay to the S$_{1/2}$ and D$_{5/2}$ states, we first
shelve the population in the D$_{3/2}$ state by a 397~nm laser
pulse (step 1). Next we empty the population in this state by a
pulse of 850~nm light (step 2). After $N$ of these cycles we
measure the population in the D$_{5/2}$ state by fluorescence
detection. (b) To measure another branching ratio, we first apply
a pulse of 393~nm light, to populate the D-manifold via the
P$_{3/2}$ state (step 1). Next we perform $N$ cycles consisting of
a pulse of 854~nm light to empty the population in the D$_{5/2}$
(step 2), followed again by a pulse of 393~nm light (step 1).
Again, after $N$ such cycles, a fluorescence measurements detects
the population in the D$_{5/2}$ state. The combination of
measurements (a) and (b) provides enough information to
unambiguously determine the branching fractions $p_j$.}
\label{fig_branchingratio}
\end{figure}

A partial level scheme of Ca$^+$ showing its five lowest energy
levels is depicted in Fig.~\ref{fig_levelscheme_Ca40}. The
P$_{3/2}$ state decays into the three states S$_{1/2}$, D$_{3/2}$,
and D$_{5/2}$, the D-states being metastable with a lifetime of
about 1.2~s \cite{Barton:2000,Kreuter:2005}. The branching
fractions $p_j$, i.~e.~ the relative strengths of the decay
processes, will be labelled by $j$, $j$ being the angular momentum
quantum number of the state into which the ion decays
($j\in\{\frac{1}{2},\frac{3}{2},\frac{5}{2}\}$).
By definition, $p_j$ fulfil the normalization condition
$p_{1/2}+p_{3/2}+p_{5/2}=1$. In this situation, a branching ratio
is conveniently measured by (i) preparing the ion in one of the
lower states, (ii) optically pumping it via the P$_{3/2}$ to the
other two states, and (iii) detecting the state populations in
these states \cite{Kurz:2008}. This scheme is illustrated by
Fig.~\ref{fig_branchingratio}(a).
An ion, initially in S$_{1/2}$, is prepared in D$_{3/2}$ by
optical pumping with light at 397~nm (step 1). Then, the state is
completely emptied by a pulse of light at 850~nm exciting the ion
to the P$_{3/2}$ state so that the ion decays into either of the
states S$_{1/2}$ and D$_{5/2}$ (step 2). Finally, a quantum state
measurement reveals the state into which the ion decayed. By
repeating this elementary sequence $M$ times and averaging over
the measurement outcomes, an estimate $\hat{r}$ of the decay
probability into D$_{5/2}$
\[
r=p_{5/2}/(p_{1/2}+p_{5/2})
\]
is obtained with a statistical error
$\sigma_{\hat{r}}=\sqrt{r(1-r)/M}$ set by quantum projection
noise. Note that in this measurement the branching fraction
$p_{5/2}$ is normalized by the factor $p_{1/2}+p_{5/2}$ to account
for the fact that population decaying back into D$_{3/2}$ is
immediately pumped back by the 850~nm laser pulse. In Ca$^+$,
earlier experiments \cite{Gallagher:1967} had shown that the
likelihood of a decay into D$_{5/2}$ was rather small ($r\approx
0.05$). For $r\ll 1$, the accuracy of the measurement is improved
by repeating steps 1 and 2 many times before performing the state
measurement. In this way, population is shuffled back and forth
between S$_{1/2}$ and D$_{3/2}$, with a growing fraction
accumulating in D$_{5/2}$. The population in this state after
applying $N$ cycles is given by
\begin{equation}
r_N=1-(1-r)^N. \label{eq_decay1}
\end{equation}
Resolving this equation for $r$ yields an estimate
$\hat{r}=1-\sqrt[N]{1-\hat{r}_N}$ for the decay probability. A
short calculation shows that the uncertainty of $\hat{r}$ can be
expressed as
\begin{equation}
\sigma_{\hat{r}}(N)=\frac{1-\hat{r}}{\sqrt{M}N}\sqrt{(1-\hat{r})^{-N}-1},
\end{equation}
where it was assumed that the measurement of $r_N$ is
quantum-limited in precision. Minimization of $\sigma_{\hat{r}}$
as a function of $N$ results in the optimum number of cycles given
by
\begin{equation}
N^\ast=\frac{x^\ast}{-\log{(1-r)}},
\end{equation}
with the constant $x^\ast\approx 1.594$ minimizing the function
$f(x)=\sqrt{e^{x}-1}/x$. For the optimum number of measurements
$N^\ast$, we have $r_N(N^\ast)=1-e^{-x^\ast}\approx 0.8$ and the
measurement uncertainty is reduced compared to the single-cycle
experiment by
\[
\frac{\sigma_{\hat{r}}(N^\ast)}{\sigma_{\hat{r}}(N=1)}=-\sqrt{\frac{1-r}{r}}\log(1-r)f(x^\ast)\approx
1.24\sqrt{r},
\]
the approximation being valid in the case $r\ll 1$. For $r=0.05$,
the optimum cycle number is given by $N^\ast=31$. The reduction in
the number of measurements needed to reach a certain level of
precision directly translates into a reduction of overall
measurement time as the duration of a single experiment is
dominated by the time required for cooling the ion and measuring
its quantum state.

A second type of experiment detecting another branching ratio is
needed for determining the branching fractions $p_j$. Towards this
end, the scheme shown in Fig.~\ref{fig_branchingratio}(b) is
employed: An ion in state S$_{1/2}$ is excited to the P$_{3/2}$
state and subsequently decays into states D$_{3/2}$ and D$_{5/2}$
(step 1). Afterwards, measurement of the D$_{5/2}$ state
population reveals the probability
\[
s=p_{3/2}/(p_{5/2}+p_{3/2})
\]
of a decay into D$_{3/2}$.

Again, the measurement accuracy can be increased by shuffling
population between the states S$_{1/2}$ and D$_{5/2}$. For this,
in a second step, the D$_{5/2}$ state is emptied by light at 854~nm. In
this step the probability of decay into D$_{3/2}$ is given by
\[
q =p_{3/2}/(p_{1/2}+p_{3/2}).
\]
Now, the measurement prescription is to perform step 1, followed
by $N$ cycles consisting of step 2 and step 1, and to measure the
D$_{5/2}$ state population given by
\begin{equation}
\label{eq_decay2} s_N=(1-s)\left\{(1-q)(1-s)\right\}^N.
\end{equation}

In combination with the normalization condition\linebreak $\sum_k
p_k=1$, the two measurements are sufficient for unambiguously
determining the branching fractions $p_j$.

\section{Setup}
\label{sec:setup}

In the following, a description of the experimental setup will be
given. A more detailed account can be found in
ref.~\cite{Benhelm:2008}. A single $^{40}$Ca$^+$ ion is trapped in
a linear Paul trap with trap frequencies of 1.24~MHz and 3~MHz in
the axial and radial direction, respectively. The ion is
Doppler-cooled by light at 397~nm on the
S$_{1/2}\leftrightarrow\mbox{P}_{1/2}$ transition. Light at 866~nm
repumps the population lost to the D$_{3/2}$ state back into the
cooling cycle. To prevent coherent population trapping in
superpositions of Zeeman states, a magnetic field of about 3~G is
applied to lift the degeneracy of the Zeeman states.

Commercially available diode lasers are used for generating
narrow-band laser light at wavelengths of 850~nm, 854~nm and
866~nm. Laser light at 393~nm and 397~nm is derived from frequency
doubled diode lasers emitting at 793~nm and 786~nm, respectively.
Furthermore, a titanium sapphire laser locked to a high finesse
cavity and operating at a wavelength of 729~nm is available for
coherent excitation of the S$_{1/2}\leftrightarrow\mbox{D}_{5/2}$
transition.

\begin{figure}
\includegraphics[width=90mm]{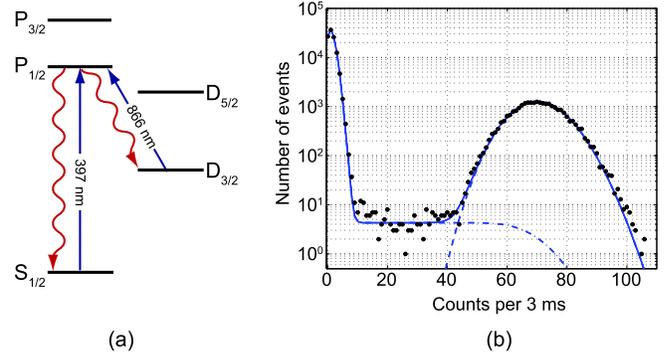} 
\caption{Quantum state detection in Ca$^+$. (a) The ion is excited
by laser light on the S$_{1/2}\leftrightarrow\mbox{P}_{1/2}$ and
D$_{3/2}\leftrightarrow\mbox{P}_{1/2}$ transitions for a few
milliseconds. Fluorescence will be detected if the ion is in
either the S$_{1/2}$ or D$_{3/2}$ state. An ion in the D$_{5/2}$
state will not fluoresce. To achieve an efficient quantum state
detection, the fluorescence detection needs to be short compared
to the lifetime of the metastable D$_{5/2}$ state but long enough
to provide a clear separation between the photon count
distributions of the `dark' and the `bright' states. (b) Photon
count distribution constructed from $10^5$ experiments. In the
bright state to the right, the ion scatters on average 70 photons
detected by the photomultiplier within the detection time of 3~ms.
In the dark state to the left, on average 1.4 background photons
are detected due to scattering of laser light from the trap
electrodes. From the fit, we infer that the ion was in the
D$_{5/2}$ with a probability of 0.803(1).} \label{fig_detection}
\end{figure}

The fluorescence of the ion emitted on the
S$_{1/2}\leftrightarrow\mbox{P}_{1/2}$ transition is detected on
an electron-multiplying CCD camera and a photomultiplier. For
quantum state detection, the ion is excited on the
S$_{1/2}\leftrightarrow\mbox{P}_{1/2}$ transition and the
fluorescence light is collected on a photo multiplier for 3-5~ms.
An average photon count rate of 23~kcounts/s is recorded when the
ion is in either of the `bright' S$_{1/2}$ and D$_{3/2}$ states.
If the ion is in the `dark' D$_{5/2}$ state, an average of
0.5~kcounts/s is detected, caused by scattering of the laser off
the trap electrodes. This scheme allows for a discrimination
between the quantum states with near-unit detection efficiency.
The detected photon counts per measurement have a Poissonian
distribution around the average value, as shown in
Fig.~\ref{fig_detection}. When the ion is in the D$_{5/2}$-state
however, the distribution is slightly modified due to the
possibility of spontaneous decay during the detection interval. As
a consequence, the Poisson distribution acquires a long tail
towards the average value of counts for the S-state. To determine
the probability of being in the bright or dark state at the end of
an experiment, we repeat the experiment $M$ times and record the
resulting photon counts. Then, a maximum likelihood method is used
to determine the probability $p$ for the ion of being in the dark
state. For each single experiment, we compute the probability of
observing the detected number of photon counts given $p$ and
multiply these probabilities for all experiments to obtain an
overall probability $\Pi$. The probability $p_D$ determined by the
maximum likelihood estimate then corresponds to the value of $p$
that maximizes the probability $\Pi$.

In order to discriminate between the S$_{1/2}$ and the D$_{3/2}$
state, the S$_{1/2}$ state population is coherently transferred to
the D$_{5/2}$ state prior to state detection by a laser exciting
the S$_{1/2}\leftrightarrow\mbox{D}_{5/2}$ quadrupole transition.
For this, either resonant laser $\pi$-pulses or rapid adiabatic
passages~\cite{Wunderlich:2007} are used connecting pairs of
Zeeman states in the ground and the metastable state. This way, we
achieve a transfer probability of better than 99.5\%.

\section{Measurement results}
\begin{figure}
\includegraphics[width=70mm]{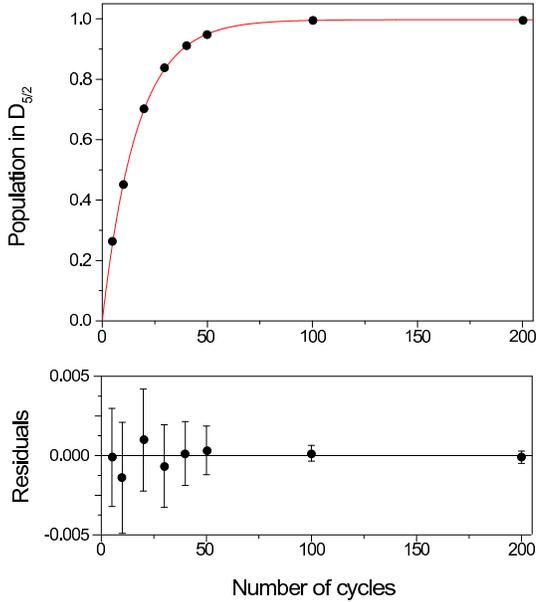} 
\caption{ Population in the D$_{5/2}$ state as a function of the
number of cycles $N$ as depicted in
Fig.~\ref{fig_branchingratio}(a). Every data point consists of at
least 2$\times$10$^4$ measurements to determine the average
population in D$_{5/2}$. The data points are jointly fitted with
the measurement results presented in Fig.~\ref{fig_decay2} to
obtain the branching fractions $p_j$ as described in the main
text. The fitted curve is given by
eq.~(\ref{eq:populations_397_850}). The lower plot shows the
residuals with errors bars given by the statistical uncertainty
due to the finite number of measurements.} \label{fig_decay1}
\end{figure}
For the experimental determination of the P$_{3/2}$ state's
branching fractions, a single ion is loaded into the trap. The
first experiment implements the scheme depicted in
Fig.~\ref{fig_branchingratio}(a). We start by measuring the time
it takes to completely empty the S$_{1/2}$ state by light at
397~nm. For this, a laser pulse of variable length is applied to
an ion prepared in state S$_{1/2}$, and subsequently the
population transferred to the D$_{3/2}$ state is detected by
shelving the remaining S$_{1/2}$ state population in the D$_{5/2}$
level. Fitting an exponentially decaying function to the S$_{1/2}$
state population, we find a $1/e$~time of 1~$\mu$s. In a similar
measurement, we determined the decay time of population in
D$_{3/2}$ to be 16~$\mu$s when emptying the state by light at
850~nm. In a next step, we implement a population shuffling cycle
by switching on the laser at 397~nm for 20~$\mu$s before turning
on the laser at 850~nm for a duration of 100~$\mu$s. We repeat
this cycle $N$ times to slowly accumulate all the population in
the D$_{5/2}$ state. Fig.~\ref{fig_decay1} shows the D$_{5/2}$
state population for cycle numbers ranging from 5 to 200 by
repeating the experiment for each value of $N$ between 20000 and
23000 times. The experimental results are fitted by slightly
modifying eq.~(\ref{eq_decay1}) to
\begin{equation}
\label{eq:populations_397_850} r_{N,exp}=A(1-(1-r)^N)
\end{equation}
by introducing a scaling factor $A$ to account for the fact that
even after 200 pumping cycles the measurement finds 0.3\% of the
population to be outside the D$_{5/2}$ state. This effect is
caused by the finite lifetime of the metastable states as will be
discussed in subsection \ref{sec:errors}.
\begin{figure}
\includegraphics[width=70mm]{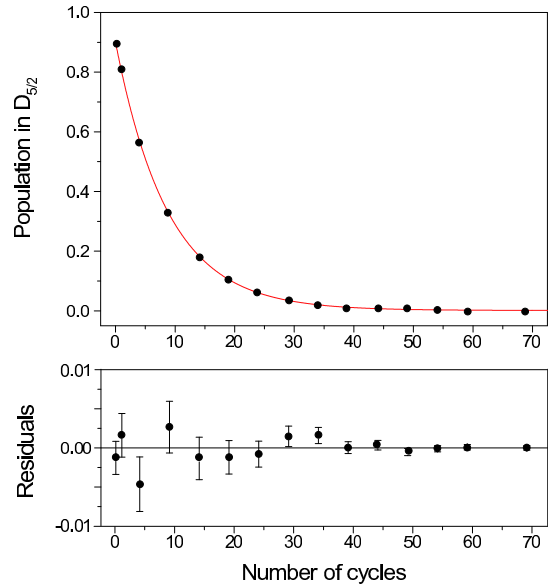} 
\caption{Population in the D$_{5/2}$ state as a function of the
number of cycles $N$ as depicted in
Fig.~\ref{fig_branchingratio}(b). The lower plot shows the
residuals after fitting the data. The fitted curve is given by
eq.~(\ref{eq:populations_393_854}).} \label{fig_decay2}
\end{figure}

A similar procedure is applied for implementing the scheme
depicted in Fig.~\ref{fig_branchingratio}(b). The $1/e$ pumping
times for emptying the states S$_{1/2}$ and D$_{5/2}$ by light at
393~nm and 854~nm are determined to be 2~$\mu$s each. Based on
this measurement, a pumping cycle - transferring population from
D$_{5/2}$ to S$_{1/2}$ and back again- was chosen that consisted
of a laser pulse at 854~nm of 20~$\mu$s duration, followed by
another 20~$\mu$s pulse at 393~nm. Again, we measure the
population remaining in the D$_{5/2}$ state after $N$ pumping
cycles, $N$ varying between 0 and 69. Fig.~\ref{fig_decay2} shows
the experimental results obtained by repeating the experiment
20000 times for each value of $N$. We fit the results by modifying
eq.~(\ref{eq_decay2}) to
\begin{equation}
 s_{N,exp}=(1-s)\left\{(1-q)(1-s)\right\}^N +
 B\label{eq:populations_393_854}.
\end{equation}
Here, the offset $B$ accounts for 0.03\% of the population
seemingly remaining in the D$_{5/2}$ state. The error bars
appearing in Fig.~\ref{fig_decay1} and Fig.~\ref{fig_decay2} are
of statistical nature due to the finite number $N$ of experiments.

The parameters $r$, $s$, $q$ appearing in equations
(\ref{eq:populations_397_850}) and (\ref{eq:populations_393_854})
can be expressed by $p_{3/2}$ and $p_{5/2}$. To determine the
branching fractions $p_j$, we use a weighted least-square fit to
jointly fit both decay curves by the set of parameter $\{p_{3/2},
p_{5/2},A,B\}$, assuming that the measurement error is given by
quantum projection noise. With this procedure, we find
$p_{1/2}=0.9347(3)$, $p_{3/2}=0.00661(4)$, $p_{5/2}=0.0587(2)$,
and $A=0.9971(3)$, $B=0.0013(2)$. The error bars are obtained from
a Monte-Carlo bootstrapping technique \cite{Efron:1986}. We
randomly vary the experimentally observed D$_{5/2}$ populations
$p$ by shifting their values assuming a Gaussian distribution with
variance $p(1-p)/M$ and fit the resulting data set. Repeating this
procedure 1000 times, the spread of the simulated fit parameters
provides us with an error estimate for the branching fractions.

\section{Discussion}
\begin{table*}
\centering \caption{\label{tab:1} Branching fractions of the decay
of the P$_{3/2}$ state as found in the present work compared to
various theoretical values from the literature. Ref.
\cite{Mitroy:2007} gives the relative decay strengths of the state
$4p$ into $4s$ and $3d$. The last two columns give the Einstein
coefficients $A_{fi}$ measured from the branching fractions in
combination with the lifetime measurement of ref.~\cite{Jin:1993}
and the theoretical prediction of ref.~\cite{Arora:2007}.
}
\begin{tabular*}{0.9\textwidth}{llllllll}
\hline\noalign{\smallskip}
Transition & \hspace{2cm} Branching fractions\hspace{-10cm} \rule{0ex}{1ex}& & & & & \hspace{1cm} $A_{fi}\times 10^6 s^{-1}$ \hspace{-1cm}  &\\
 & Present work & ref.~\cite{Guet:1991} & ref.~\cite{Liaw:1995} &  ref.~\cite{Mitroy:2007} & ref.~\cite{Arora:2007}\hspace{1cm} & Present work+\cite{Jin:1993} & ref. \cite{Arora:2007}\\
\noalign{\smallskip}\hline\noalign{\smallskip}
4P$_{3/2} \leftrightarrow$ 4S$_{1/2}$ & 0.9347(3)  & 0.9381 & 0.9354 & 0.9357 & 0.9340 & 135.0(4) & 139.7 \\
4P$_{3/2} \leftrightarrow$ 3D$_{3/2}$ & 0.00661(4)  & 0.00628 & 0.00649 & & 0.00667 & 0.955(6)& 0.997 \\
4P$_{3/2} \leftrightarrow$ 3D$_{5/2}$ & 0.0587(2)  & 0.0556 & 0.0581 & \raisebox{1.5ex}[-1.5ex]{0.0643} & 0.0593 & 8.48(4)  & 8.877\\
\noalign{\smallskip}\hline
\end{tabular*}
\end{table*}

\subsection{Statistical errors}
\label{sec:staterrors}

In the regression analysis used for determining the branching
fractions, we assume the errors to be quantum-noise limited. This
means that we consider the outcome of an individual experiment to
be given by a random variable $X_i$, $i=1,\ldots,M$, yielding the
outcome `1' with probability $p$ and the outcome `0' with
probability $1-p$. For a set of $M$ experiments, we then assume
the standard deviation of $X=\frac{1}{M}\sum_{i=1}^M X_i$ to be
given by $\sigma_p=\sqrt{p(1-p)/M}$. If, however, due to
experimental imperfections the probability $p$ slightly changes
over the course of time needed for carrying out the $M$
experiments, this assumption would be violated. We therefore
checked the validity of our hypothesis by subdividing our data
into $K$ sets S$_k$, $k=1,\ldots,K$, each containing $M/K$
consecutively taken experiments described by the random variables
$\{X_{j_k+1},\ldots,X_{j_k+M/K}\}$. For each set S$_k$, we compute
its mean value\linebreak $p^{(k)}=\frac{K}{M}\sum_{i\in
S_k}\langle X_i\rangle$ and its deviation $p^{(k)}\!-\!p$ from the
average probability. We repeat this procedure for different values
of $K$ ranging from 2 to 1000. A comparison of the distribution of
$p^{(k)}\!-\!p$ with the one expected for truly random variable
all having the same probability $p$ leads us to conclude that the
assumption of quantum-limited errors is well-justified.

\subsection{Systematic errors}
\label{sec:errors}

The method of measuring the branching fractions by repeated
optical pumping assumes that the optical pumping steps are
perfect. If, however, a small fraction $\epsilon_{1,2}$ is not
pumped out in the two pumping steps of the cycle, the measured
decay rates will be smaller than for perfect optical pumping. This
systematic effect is taken into account by defining an effective
cycle number $N_{eff}=N(1-\epsilon_1-\epsilon_2)$ that replaces
the cycle number $N$ in equations (\ref{eq:populations_397_850})
and (\ref{eq:populations_393_854}). This correction was actually
applied in the data evaluation of the first experiment as it
turned out that the laser pulse at 850~nm of 100~$\mu$s duration
was too short to completely empty the D$_{3/2}$ state
($\epsilon_2=0.3$\% of the population were left behind).

The finite lifetime $\tau_D=1.2$~s of the metastable D-states also
has a small effect on the measured branching ratios. The quantum
state detection method we use is not affected by spontaneous decay
processes occurring during the detection time. Spontaneous decay
during the pumping cycle, however, slightly modifies the results.
This is best illustrated by the pumping scheme shown in
Fig.~\ref{fig_branchingratio}(a). Spontaneous decay of the
D$_{3/2}$ state is negligible since the state is emptied by light
at 850~nm in 16$\mu$s, i.e. with a rate roughly 10$^4$ times
bigger than the spontaneous decay rate. The situation is different
for the D$_{5/2}$ state which is populated by a rate
$R_a=r\tau_{cycle}^{-1}$ where $\tau_{cycle}=120\,\mu$s is the
duration of a single pumping cycle. Here, spontaneous decay gives
rise to a steady-state D$_{5/2}$ state population
$p_{D,\infty}=1-R_a\tau_D\approx 99.8\%$ in the limit of an
infinite number of pumping cycles which is consistent with the
measured value of the fit multiplier $A$. In addition to changing
the steady state solution, the additional decay process also
decreases the time required for reaching the equilibrium (see
eq.~(\ref{eq:populations_397_850})). This effect was taken into
account in the analysis of the branching fractions stated in
Tab.~\ref{tab:1}.

A similar argument can be made for spontaneous decay affecting the
pumping scheme of Fig.~\ref{fig_branchingratio}(b). Here, the
D$_{3/2}$ is populated with a rate $R_b\approx s\tau_{cycle}^{-1}$
which needs to be compared to the spontaneous decay rate
$1/\tau_D$. Due to this effect, we expect to find a steady-state
population of about 0.04\% that is not in D$_{3/2}$. Here, the
effect is smaller since $r<s$ and the duration of the pumping
cycle was about five time shorter than in the other experiment.
The correction to the measured rate is also about 0.04\% and does
not significantly alter our measurement. Before performing the
branching ratio measurements, we had checked that the lifetimes of
the metastable states were not significantly shortened by the
lasers at 850 and 854~nm which could occur either by a broad-band
frequency background of the diode lasers producing the light or by
imperfectly switched off laser beams.

In general, any additional mechanism that leads to an exchange of
population between the atomic states will modify the steady-state
population obtained in the limit of large $N$ and shorten the time
scale required for reaching the steady-state. Spin-changing
collisions leading to transitions between the D$_{3/2}$ and the
D$_{5/2}$ state are of no importance in our measurements as they
occur at a rate that is smaller than $10^{-2}s^{-1}$.

\subsection{Comparison with other measurements and calculations}

The branching fractions obtained from fitting the two decay
measurement can be compared to previous measurements and
theoretical calculations. By evaluating the ratio
$p_{3/2}/(p_{1/2}+p_{5/2})$, we find $A_{(\rm P_{3/2}- \rm S_{1/2}})/\sum_J
A_{(\rm P_{3/2}-D_J)}=14.31(5)$ for the branching ratio of the decay
into the S- and D-states. This is a bit lower than the value of
17.6(2.0) measured by Gallagher~\cite{Gallagher:1967} in a
discharge experiment in 1967. A comparison with theoretical
calculations \cite{Guet:1991,Liaw:1995,Mitroy:2007,Arora:2007} is
given in Tab.~\ref{tab:1}, showing a good agreement between
experiment and theory.

When combining the branching ratio measurements with an
experimental determination \cite{Jin:1993} of the P$_{3/2}$
state's lifetime, the relative decay strengths can be converted
into Einstein $A$ coefficients which are also given in Tab.
\ref{tab:1}. We also list the theoretical predictions of
ref.~\cite{Arora:2007} which are all about 4\% higher, a
discrepancy much bigger than for the relative decay strengths.
From this point of view, it might be of interest to use a single
trapped ion for a measurement of its excited state lifetime
\cite{Moehring:2006} or its absolute oscillator strengths.

\subsection{Measuring Einstein coefficients in a single ion}
A large variety of methods exists for measuring transition matrix
elements in ensembles of neutral atoms or ions \cite{Huber:1986}.
In view of the discrepancy between the Einstein coefficients
calculated from experimental observations and predicted by theory,
we would like to propose yet another technique capable of directly
measuring Einstein coefficients in a single ion. We will discuss
the technique for the case of the decay rate of the P$_{1/2}
\leftrightarrow\mbox{D}_{3/2}$ transition.

We start with an ion initially prepared in the S$_{1/2}$ state. A
narrow-band laser off-resonantly exciting the\linebreak
S$_{1/2}\leftrightarrow\mbox{P}_{1/2}$ transition pumps the ion
into the D$_{3/2}$ state. If the detuning $\Delta$ of the laser is
much bigger than the inverse of the excited state lifetime, the
pumping rate is given by $R=A_{({\rm P}_{1/2}-{\rm
D}_{3/2})}\rho_{\,{\rm P}_{1/2}}$ where $\rho_{\,{\rm
P}_{1/2}}=\Omega^2/(4\Delta^2)$ is the excited state population
and $\Omega$ the Rabi frequency of the laser exciting the
transition. The Rabi frequency is related to the ac-Stark shift
$\delta_{ac}=\Omega^2/(4|\Delta|)$ experienced by the S$_{1/2}$
state due to the presence of the off-resonant light. In this way,
the Einstein coefficient can by expressed by
\[
A_{({\rm P}_{1/2}-{\rm D}_{3/2})}=R\frac{\Delta}{\delta_{ac}}.
\]
A measurement of the pumping rate could be performed in a similar
way as the experiments presented in this paper by shelving the
S$_{1/2}$ population in D$_{5/2}$ prior to quantum state
detection. For the determination of the light shift $\delta_{ac}$,
a spin echo experiment could be performed on the
S$_{1/2}\leftrightarrow\mbox{D}_{5/2}$ transition, measuring the
shift of the transition frequency by the light pumping the ion.
Measurements of this type have recently been
shown~\cite{Roos:2008} to be able to resolve level shifts as small
as 1~Hz. To give an example, we assume
$\Delta=(2\pi)\,10^{10}$~s$^{-1}$ and adjust the Rabi frequency to
get $R=10^2$~s$^{-1}$. Because of $A({\rm P}_{1/2}-{\rm
D}_{3/2})\approx 10^7\,$s$^{-1}$, the light shift will be
$\delta_{ac}=(2\pi)\,10^5$~s$^{-1}$. To measure the Einstein
coefficient to better than 1\% would require a measurement of the
light shift with a precision of better than 1~kHz within a
measurement time somewhat shorter than $1/R=10$~ms which appears
to be feasible.

For a proper treatment, the Zeeman sublevels of the atomic levels
need to be taken into account. Fortunately, in the case of the
S$_{1/2}\leftrightarrow\mbox{P}_{1/2}$ transition, the pumping
rate and the induced light shift are independent of the
polarization of the exciting laser as long as it is linearly
polarized. In this way, the Einstein coefficient could be
independently measured with high precision to complement the
branching ratio measurements presented in this paper.

\subsection{Conclusion}
Our experiments demonstrate that single trapped ions are perfectly
suited for a precision measurement of branching fractions. As
compared to experiments based on a single pumping step, the
technique of repeated optical pumping offers two important
advantages. Firstly, it reduces the number of measurements
required for obtaining a given level of accuracy in the case where
the transition probabilities to the lower-lying states are very
different. Secondly, monitoring the population transfer as a
function of the number of the pumping cycle provides a means to
detect possible sources of errors like atomic state changes
induced by sources other than the lasers used for optical pumping.

The pumping technique used in this paper requires an atom with a
tripod level structure, i.~e. an upper state decaying into three
lower-lying states. As this level structure exists not only in
Ca$^+$ but also in many other isotopes, the method should be
widely applicable.

\begin{acknowledgement}
We gratefully acknowledge the support of the European network
SCALA, the Institut f{\"u}r Quanteninformation GmbH and IARPA.
R.~G. acknowledges funding of the Marie-Curie program of the
European Union (grant number PIEF-GA-2008-220105). C.~F.~R. would
like to thank J.~Mitroy and M.~Safronova for useful discussions.
\end{acknowledgement}

%

\end{document}